\documentclass[%
 aip,
 jmp,%
 amsmath,amssymb,
 manuscript,%
]{revtex4-1}

\usepackage{graphicx}
\usepackage{dcolumn}
\usepackage{bm}
\usepackage{booktabs}

\begin{document}


\title{On the feasibility of \textit{ab initio} electronic structure calculations for Cu using a single \textit{s} orbital basis}

\author{Ganesh Hegde}
	\email{ganesh.h@ssi.samsung.com}
\author{R. Chris Bowen}%
\affiliation{Advanced Logic Lab, Samsung Semiconductor Inc., Austin, TX 78754}
\date{\today}
\begin{abstract}
The accuracy of a single $s$-orbital representation of Cu towards enabling multi-thousand atom \textit{ab initio} calculations of electronic structure is evaluated in this work. If an electrostatic compensation charge of 0.3 electrons per atom is used in this basis representation of copper, the electronic transmission in bulk and nanocrystalline Cu compares accurately to that obtained with a Double Zeta Polarized basis set. The use of this representation is analogous to the use of single band effective mass representation for semiconductor electronic structure. With a basis of just one $s$-orbital per Cu atom, the representation is extremely computationally efficient and can be used to provide much needed \textit{ab initio} insight into electronic transport in nanocrystalline Cu interconnects at realistic dimensions.
\end{abstract}

\maketitle
Copper is the current and and projected metal of choice for back-end-of-line (BEOL) interconnects used in semiconductor logic and memory technology. It has been well known that resistivity trend for copper is strongly non-linear and has been increasing rapidly with a reduction in interconnect width. A number of semi-empirical models have been formulated to explain this phenomenon. Broadly speaking, these models can be categorized as variants of the Fuchs-Sondheimer model (F-S) \cite{fuchs1938conductivity, sondheimer1952mean} of surface scattering-induced increase in resistivity or the Mayadas-Shatzkes (M-S) model \cite{mayadas1970electrical} of grain-boundary scattering-induced increase in resistivity.

While these models provide important insights into electron transport in metals, they are fit \textit{a posteriori} to experimental data with empirical parameters such as average reflectivity and specularity. Such parameteric models can be fit with a non-unique set of parameters and consequently provide very little predictive insight from a materials design perspective. For instance, new experimental data on nanocrystalline copper requires a recalibration of the model to account for the new data. Often times, such empirical models can lead to conflicting physical insight. For instance, Graham et al. \cite{graham2010resistivity} fit the resistivity of sub-100 nm Cu interconnect lines to a purely surface scattering based model, while Steinh\"{o}gl et al.\cite{steinhogl2005comprehensive} fit the resisitivity of Cu lines in the same dimensional range to a predominantly grain boundary scattering based model.

In spite of the apparent seriousness of the resistivity runaway problem and lack of understanding of the fundamental features that govern electronic transport in nanocrystalline (polycrystalline, with nanometer-range sized grains) Cu structures, there have been very few  first principles based investigations on electron transport in nanocrystalline Cu \cite{feldman2010simulation,ke2009resistivity,timoshevskii2008influence,zhou2010ab,cesar2014calculated}. Part of the reason for such a paucity of first principles based models is the cumbersome computational requirement (memory and execution time) for convergence of first principles transport calculations for realistic Cu nanocrystalline systems. This is illustrated with a simple example - The ITRS projects the smallest damascene interconnect cross sectional area of $10\times22$ nm$^2$ for damascene Cu in the year 2025 \cite{wilson2013international}. Cu has a lattice constant of 3.61 $\AA$ at room temperature. A cross sectional sliver of a length of just 1 nm with the aforementioned area of Cu at the projected ITRS dimensions contains approximately (assuming bulk density holds even for nanocrystalline samples) 18700 atoms. If one were to simulate such a sliver using linear scaling order-N Linear Combination of Atomic Orbitals (LCAO) density functional theory (DFT) calculations with a moderately sized basis set (Double Zeta Polarized (DZP)), containing 2 $s$, 10 $d$ and 3 $p$ orbitals, one would require approximately 1.25 TB to store a full representation of this system in memory (64 bit word size) for DFT calculations. Even if one were to reduce this to a sparse representation, one would require several hundred GB since $s$ orbital interactions in Cu are long ranged. Apart from the fact that memory requirements (for full storage) for a system of size $N$ atoms and an $M$ orbital basis per atom scale as ($N \times M$)$^2$, full eigenvalue solves (required for density computation in metals) scale as O(($N \times M$)$^3$). This is in addition to the O($N \times M$) time complexity of evaluating Overlap and Hamiltonian integrals in DFT \cite{soler2002siesta}.

What is needed then, is an \textit{ab initio} method that is computationally efficient for systems containing several thousand Cu atoms while retaining an acceptable level of accuracy. In this respect, Cu possesses a unique advantage as compared to other metals such as partially $d$ filled transition metals. Cu has a valence electron configuration of 3$d^{10}$4$s^{1}$. The $d$ orbitals are spatially localized and completely filled while the $s$ orbital is only partially filled and delocalized. An obvious question then arises - is it feasible to to represent Cu's electronic structure \textit{ab initio} using only the partially filled and delocalized $s$ orbitals while retaining acceptable accuracy? If the answer were to be in the affirmative, such a method would allow tremendous computational savings, allowing the \textit{ab initio} investigation of multi-thousand atom Cu systems.

In this paper we report the results of our investigation into this question. We provide conditions for which the electronic structure of systems represented in a single $s$ orbital representation can be compared favorably with that of DZP representations, which have in turn compared favorably to experiment in previous work \cite{CuModel, CuConfinement, feldman2010simulation}. Our end goal in this investigation is evaluation of the feasibility of using a reduced basis in studying electron transport in nanocrystalline Cu structures where the scattering mean free path is significantly smaller than the phonon scattering mean free path. Accordingly, we use ballistic transmission and ballistic transmission per unit area as figures of merit for comparing electronic structure throughout this paper.

The rest of this paper is organized as follows. We first describe the computational methodology used in our investigation. This is followed by a comparison of electronic structure computed using the two sets of bases for a variety of boundary conditions and polycrystalline configurations. We then conclude with a discussion on the relative accuracy of the method an its applicability in transport calculations on multi-thousand atom systems.

\begin{figure*}
	\centering
		\includegraphics[width=1.00\textwidth]{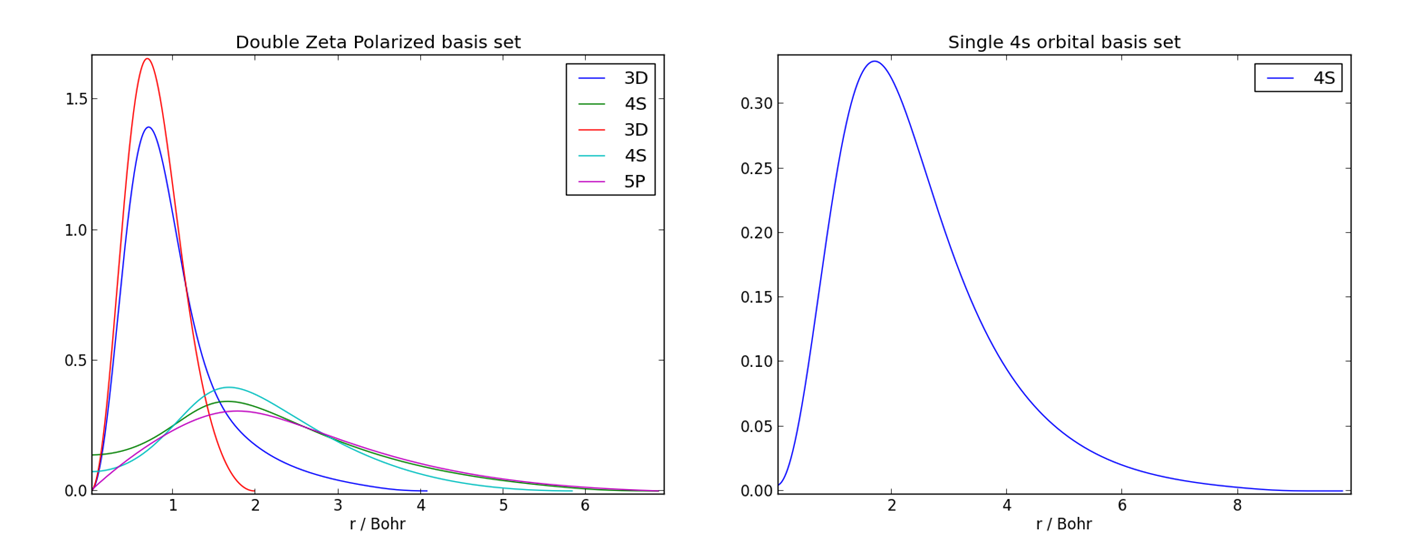}
	\caption{Radial part of the Numerical Orbitals used in the DFT calculations. On the left is a Double Zeta Polarized (DZP) basis set containing $s$, $p$ and $d$ orbitals. On the right is a single orbital basis set containing just one $s$ orbital. }
	\label{fig:DZP_Sorbital_bases}
\end{figure*}

We performed first principles LCAO DFT calculations using the Atomistix Toolkit Package (ATK) \cite{atk} within the Local Density Approximation (LDA) and the exchange correlation functional of Perdew And Zunger (PZ) \cite{perdew1981self}. The reference basis was a numerical DZP basis consisting of $s$ orbitals, $d$ orbitals and $p$ orbitals. The radial part of this basis is shown in figure \ref{fig:DZP_Sorbital_bases}. The radial part of the single $4s$ orbital numerical basis is also shown on the right in the same figure. 

\begin{figure*}
	\centering
		\includegraphics[width=1.00\textwidth]{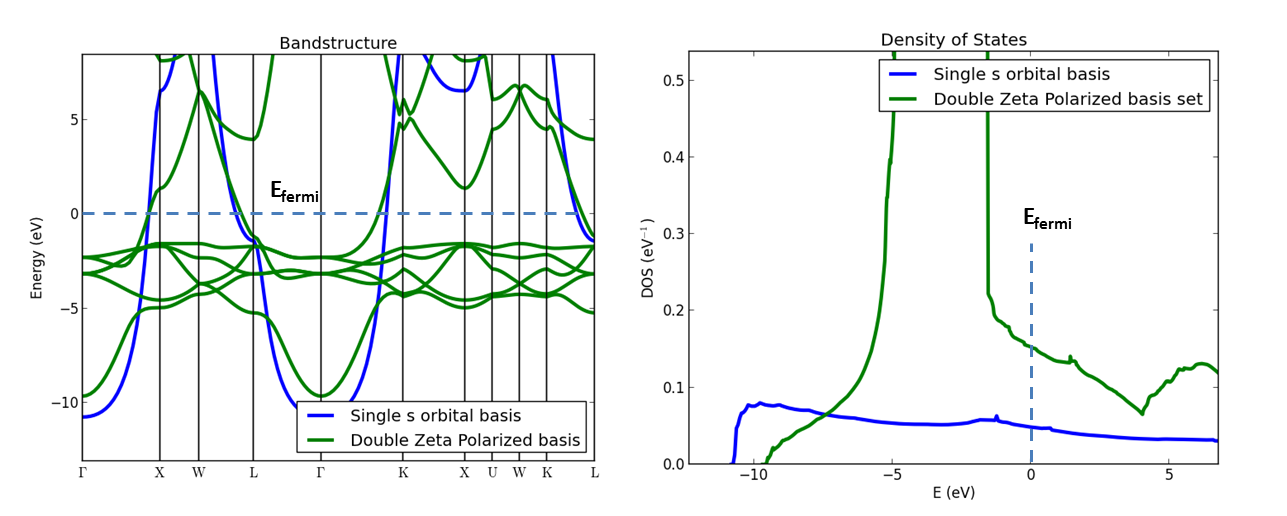}
	\caption{Comparison of band structure (left) and DOS (right) computed in LDA DFT using the two bases.}
	\label{fig:Bandstructure_and_DOS}
\end{figure*}

The band structure and density of states (DOS) computed using the two bases and bulk $k$ grid of 20$^3$ k points is shown in figure \ref{fig:Bandstructure_and_DOS}. An important feature of Cu's electronic structure is the 'neck' at the $L$ symmetry point in the Brillouin Zone. This neck is situated about 1.23 eV below the Fermi Energy in the DZP computed band structure. If the $s$ orbital basis is used as is, the neck occurs 1.3 eV above the Fermi Energy. The bands must be aligned correctly to reproduce key features of the Fermi Surface with acceptable accuracy. In order to do this, we compensated (electrostatically 'doped') the bulk Cu configuration with 0.3 electrons. This is equivalent to introducing a overall negative compensation charge of 0.3 electrons per atom in the unit cell. The resultant band structure reproduces the relative position of the $L$ neck as shown in figure \ref{fig:DZP_Sorbital_bases}.

\begin{figure*}[h]
	\centering
		\includegraphics[width=1.00\textwidth]{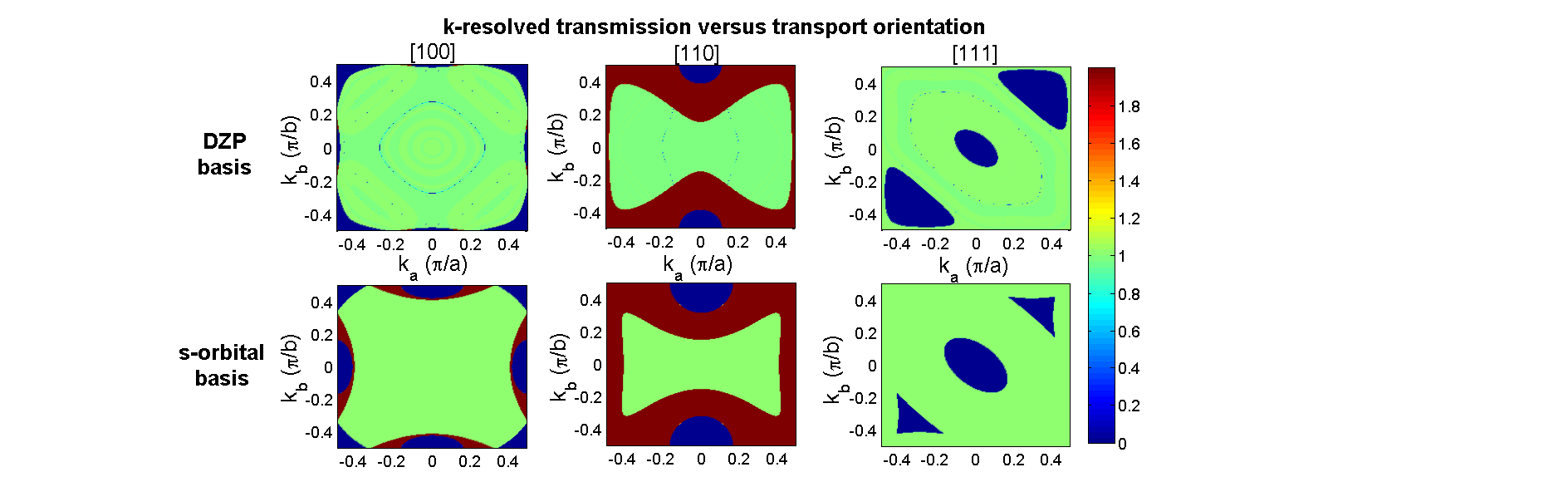}
	\caption{Transverse $k$ resolved Transmission spectra for Cu computed using the DZP basis (top) and single $s$-orbital basis (bottom)}
	\label{fig:FermiSurfaces}
\end{figure*}

It is also evident from figure \ref{fig:Bandstructure_and_DOS} that the curvature of the bands in the $s$ orbital basis is significantly higher than that computed in the DZP orbital basis. On the other hand the DOS in the $s$ orbital basis is significantly lower. Given that the net ballistic transmission is a product of DOS and average group velocity \cite{datta2005quantum}, it can be expected that the increase in group velocity is balanced out by the increase in DOS and that the ballistic transmission computed using the two bases is comparable. This is evident in the k-resolved transmission spectra computed in the Non Equilibrium Greens Function (NEGF) formalism \cite{datta2005quantum,brandbyge2002density} shown in figure \ref{fig:FermiSurfaces}.

\begin{figure*}[h]
	\centering
		\includegraphics[width=1.00\textwidth]{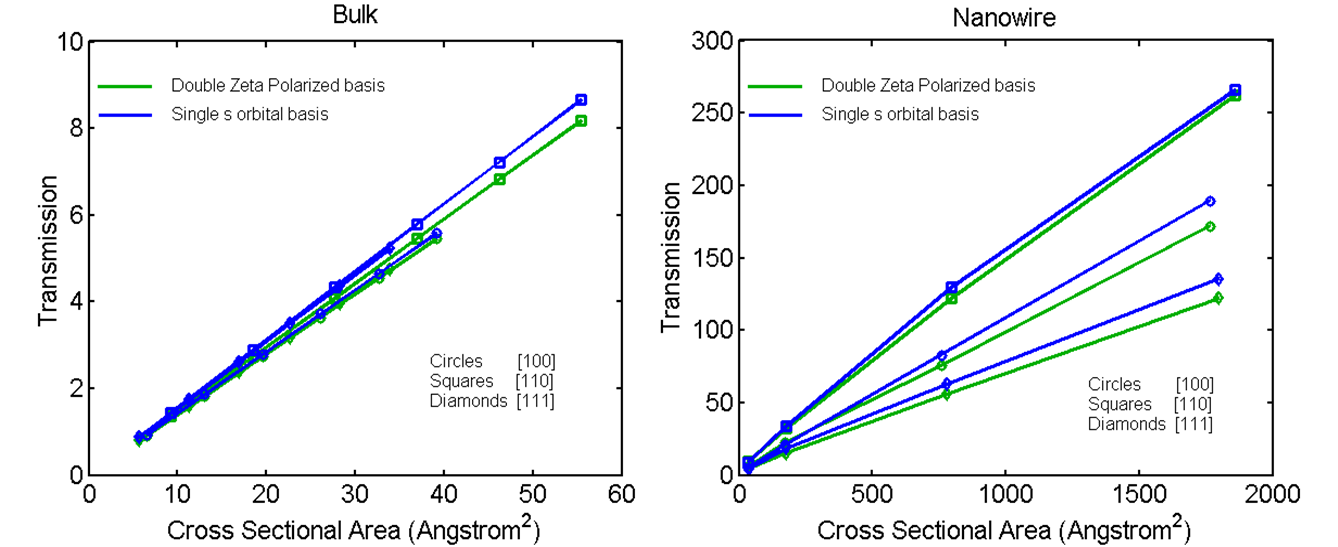}
	\caption{Transmission versus Cross Sectional Area for bulk and nanowire copper plotted for different transport orientations Computed using the $s$-orbital and DZP bases}
	\label{fig:Bulk_nw_transmission_vs_area}
\end{figure*}

The transmission spectra shown in figure \ref{fig:FermiSurfaces} were computed at the Fermi Level for the smallest unit cells in [100], [110] and [111] transport orientations. While it is evident from this figure that while the Fermi transmission spectra for the different transport orientations computed using the single $s$-orbital basis is not exactly the same as that computed using the DZP basis, the qualitative similarity between the two is quite significant considering that the $s$-orbital basis set is a severely reduced representation. In order to effect a more quantitative comparison, the transverse momentum-averaged transmission was computed for each energy point and a net transmission was obtained by Fermi-averaging the energy resolved transmission. This was then done for multiple cross sectional areas. The resultant Transmission versus Cross Sectional Area plot is shown in figure \ref{fig:Bulk_nw_transmission_vs_area}.

It is evident from the computed bulk transmission and Fermi Surfaces that the $s$-orbital model matches the DZP results quantitatively in bulk. More interesting for practical applications, however, is the behavior under confinement and when sources of disorder such as grain boundaries are present. To facilitate this comparison, Cu nanowire cross sections oriented along different orientations with a $1\times 2$ aspect ratio were constructed. The average transmission along transport direction was computed versus cross sectional area and plotted alongside the bulk values in figure \ref{fig:Bulk_nw_transmission_vs_area}. It is evident that the nanowire transmission computed using the two sets of bases matches well quantitatively. The trend in transmission mismatch between orientations for different orientations is also matched.

To compare the transmission in grain boundary structures, structures oriented along the canonical orientations [100], [110], [111] were combined and four separate boundary conditions were applied
\begin{itemize}
\item Bulk (periodic in transverse direction) with a single grain boundary and open (device) boundary conditions along transport direction.
\item Bulk with grain boundary repeated infinitely along transport direction.
\item Nanowire with a single grain boundary and open boundary conditions along transport direction.
\item Nanowire with grain boundary repeated infinitely along transport direction.
\end{itemize}
Since the number of possible configurations considering grain size and orientation distribution is potentially infinite and computationally cumbersome to compute in the DZP basis, we limit ourselves to a limited number of grain orientation distributions and a grain size of 1nm.

\begin{figure}[htbp]
	\centering
		\includegraphics[width=1.00\textwidth]{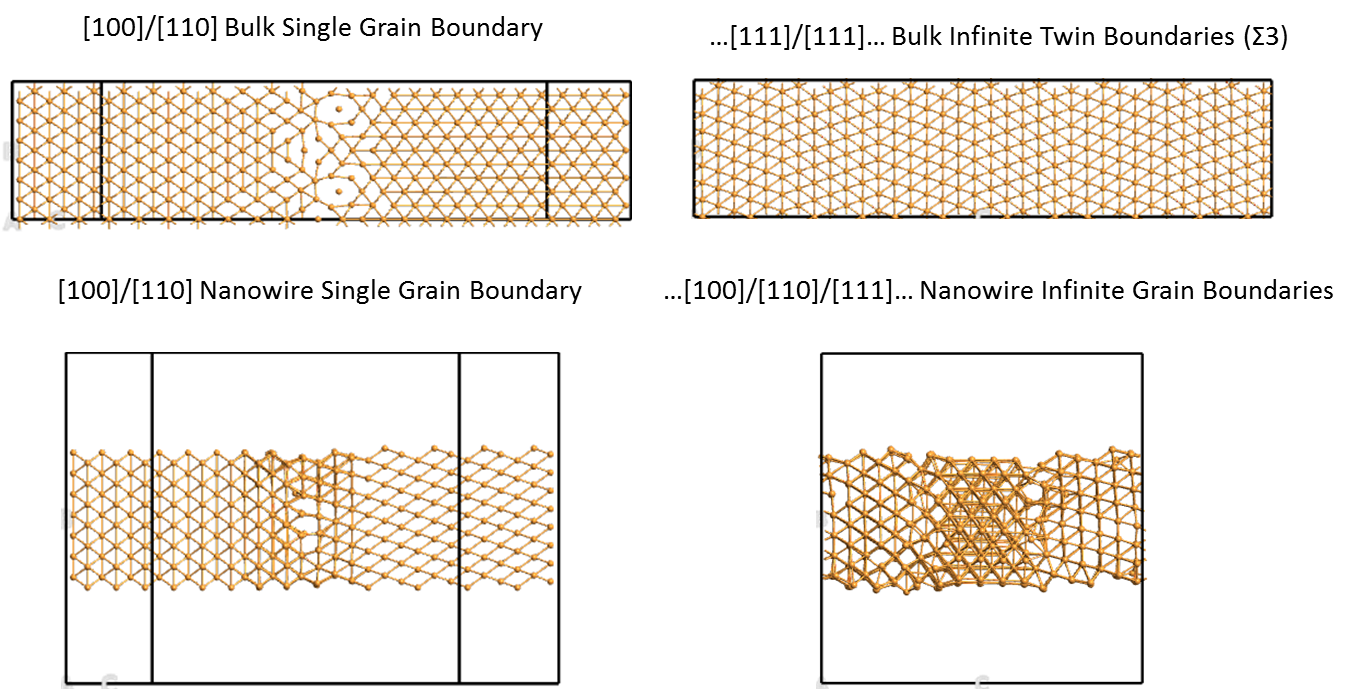}
	\caption{Examples of the grain boundary structures for which transmission was computed in the $s$-orbital and DZP bases. The black lines indicate the unit cell used in the calculation. For single grain boundary structures the area on either side of the grain boundary is repeated semi-infinitely in the transport direction.}
	\label{fig:Structures}
\end{figure}

An example of a structure for each boundary condition is shown in figure \ref{fig:Structures}. Tables \ref{tab:table1} and \ref{tab:table2} compare the transmission per unit area computed using the two sets of bases.
Of special interest is the $\Sigma$3 [111]/[111] twin boundary that has the lowest specific contact resistivity of all grain boundaries \cite{cesar2014calculated} and is the least disordered grain boundary structure in Cu.

\begin{table}[htbp]
  \centering
  \caption{Comparison of Single Grain Boundary Transmission per Cross Sectional Area for Double Zeta Polarized (DZP) and $s$-orbital basis}
		\begin{ruledtabular}
    \begin{tabular}{cccccc}
    \multicolumn{3}{c}{Bulk} & \multicolumn{3}{c}{Nanowire} \\
    Grain  Distribution & \multicolumn{2}{c}{T/A (A$^{-2}$)} & Grain Distribution & \multicolumn{2}{c}{T/A (A$^{-2}$)} \\
          & DZP   & $s$-orbital &       & DZP   & $s$-orbital \\
    \midrule \relax
    [100]/[110] & 0.119  & 0.113  & [100]/[110] & 0.092 & 0.096 \\ \relax
    [100]/[111] & 0.118  & 0.112  & [100]/[111] & 0.075 & 0.074 \\ \relax
    [110]/[111] & 0.112  & 0.101  & [110]/[111] & 0.063 & 0.070 \\ \relax
    [111]/[111] $\Sigma$3 & 0.135  & 0.145  & [111]/[111] $\Sigma$3 & 0.102 & 0.099 \\
    \end{tabular}%
		\end{ruledtabular}
  \label{tab:table1}%
\end{table}%

The results in figures \ref{fig:DZP_Sorbital_bases} through \ref{fig:Structures} and tables \ref{tab:table1} through \ref{tab:table2} show unambiguously that the transmission computed using the single $s$-orbital basis matches the qualitative trends in the transmission computed using the DZP basis, regardless of boundary condition, grain orientation distribution or structure. Quantitatively speaking, an excellent match is obtained for bulk transmission, monocrystalline and grain boundary structures. For nanowires, an extremely accurate match is obtained for monocrystalline structures. For grain boundary structures, however, the quantitative match is not as accurate even if the qualitative trend is accurate. For instance a maximum error of 36 \% is obtained for the the nanowire structures (multi grain boundary [100]/[110] case).

\begin{table}[htbp]
  \centering
  \caption{Comparison of Multiple Grain Boundary Transmission per Cross Sectional Area for Double Zeta Polarized (DZP) and $s$-orbital basis}
		\begin{ruledtabular}
    \begin{tabular}{cccccc}
    \multicolumn{3}{c}{Bulk} & \multicolumn{3}{c}{Nanowire} \\
    Grain  Distribution & \multicolumn{2}{c}{T/A (A$^{-2}$)} & Grain Distribution & \multicolumn{2}{c}{T/A (A$^{-2}$)} \\
          & DZP   & $s$-orbital &       & DZP   & $s$-orbital \\
    \midrule
    ...[100]/[110]... & 0.079  & 0.082  & ...[111]/[100]... & 0.033  & 0.021 \\
    ...[100]/[110]/[111]... & 0.070  & 0.053   & ...[100]/[111]... & 0.024  & 0.019 \\
    ...[111]/[111]... $\Sigma$3 & 0.130  & 0.140   & ...[111]/[111]... $\Sigma$3 & 0.076 & 0.083 \\
    \end{tabular}%
		\end{ruledtabular}
  \label{tab:table2}%
\end{table}%

Nevertheless, there are important reasons why this mismatch should not prevent the use of the single $s$-orbital basis in transport calculations in lieu of DZP or other 'full' basis sets. The cross sectional areas considered in this paper were limited to a maximum cross sectional area of 181 \AA$^2$ owing to difficulties in converging structures larger than 800-900 atoms using the DZP basis set. This is far smaller than current or projected interconnect dimensions. It is evident from the results above that as the surface area to volume ratio increases (as one goes towards bulk systems), the accuracy of transmission calculated in the $s$-orbital basis set improves. Additionally, it is important to note that the projected resistivity runaway trend \cite{wilson2013international} is non-linear and can be fit with exponential behavior. Even with an error bar of 20 to 40 \% and correct qualitative trends an immense amount of insight can be gained about resistivity behavior of Cu at current and future nodes. In addition, the $s$-orbital basis could be very useful in gaining relatively quick insight about the lower limits of Copper resistivity at a given dimension or the effects of surface roughness on Cu resistivity at larger cross sectional areas. An added advantage of using this basis is that there are no further adjustable parameters once the bulk bands are aligned using doping (in our case, a doping of 0.3 electron per atom was used throughout).

It should also be pointed out that the feasibility evaluation carried out above is restricted to electronic structure. This is similar in spirit to attempts at modeling semiconductor electronic structure by use of a single effective mass. Total energy calculations using a single $s$-orbital give results that do not match DZP total energies since $d$ orbitals contribute significantly to total energy in Cu.

In conclusion, we have evaluated the accuracy of a single $s$-orbital representation of Copper towards enabling multi-thousand atoms \textit{ab initio} calculations of electronic structure. We found that upon modification through doping, the $s$-orbital representation accurately reproduces key features of the bulk Fermi surface, the electronic structure of monocrystalline nanowires, bulk and nanowire single grain boundary structures and twin boundary structures. The electronic structure of multi grain boundary structures showed less comparative quantitative accuracy, but accurately matched all qualitative trends obtained through DZP calculations. The use of this representation is analogous to the use of single band effective mass representation for semiconductor electronic structure. We have found that the representation allows easy scaling to multi-thousand atom systems due to use of a reduced basis.  With a basis of just one $s$-orbital per Cu atom, the representation is extremely computationally efficient and can be used to provide much needed \textit{ab initio} insight into electronic transport in nanocrystalline Cu interconnects containing several thousand atoms.

We thank Mark Rodder for a careful reading of the manuscript and for helpful suggestions made during the course of this work.
\nocite{*}
\bibliography{bibliography}

\begin{thebibliography}{19}%
\makeatletter
\providecommand \@ifxundefined [1]{%
 \@ifx{#1\undefined}
}%
\providecommand \@ifnum [1]{%
 \ifnum #1\expandafter \@firstoftwo
 \else \expandafter \@secondoftwo
 \fi
}%
\providecommand \@ifx [1]{%
 \ifx #1\expandafter \@firstoftwo
 \else \expandafter \@secondoftwo
 \fi
}%
\providecommand \natexlab [1]{#1}%
\providecommand \enquote  [1]{``#1''}%
\providecommand \bibnamefont  [1]{#1}%
\providecommand \bibfnamefont [1]{#1}%
\providecommand \citenamefont [1]{#1}%
\providecommand \href@noop [0]{\@secondoftwo}%
\providecommand \href [0]{\begingroup \@sanitize@url \@href}%
\providecommand \@href[1]{\@@startlink{#1}\@@href}%
\providecommand \@@href[1]{\endgroup#1\@@endlink}%
\providecommand \@sanitize@url [0]{\catcode `\\12\catcode `\$12\catcode
  `\&12\catcode `\#12\catcode `\^12\catcode `\_12\catcode `\%12\relax}%
\providecommand \@@startlink[1]{}%
\providecommand \@@endlink[0]{}%
\providecommand \url  [0]{\begingroup\@sanitize@url \@url }%
\providecommand \@url [1]{\endgroup\@href {#1}{\urlprefix }}%
\providecommand \urlprefix  [0]{URL }%
\providecommand \Eprint [0]{\href }%
\providecommand \doibase [0]{http://dx.doi.org/}%
\providecommand \selectlanguage [0]{\@gobble}%
\providecommand \bibinfo  [0]{\@secondoftwo}%
\providecommand \bibfield  [0]{\@secondoftwo}%
\providecommand \translation [1]{[#1]}%
\providecommand \BibitemOpen [0]{}%
\providecommand \bibitemStop [0]{}%
\providecommand \bibitemNoStop [0]{.\EOS\space}%
\providecommand \EOS [0]{\spacefactor3000\relax}%
\providecommand \BibitemShut  [1]{\csname bibitem#1\endcsname}%
\let\auto@bib@innerbib\@empty
\bibitem [{\citenamefont {Fuchs}(1938)}]{fuchs1938conductivity}%
  \BibitemOpen
  \bibfield  {author} {\bibinfo {author} {\bibfnamefont {K.}~\bibnamefont
  {Fuchs}},\ }\bibfield  {title} {\enquote {\bibinfo {title} {The conductivity
  of thin metallic films according to the electron theory of metals},}\ }in\
  \href@noop {} {\emph {\bibinfo {booktitle} {Mathematical Proceedings of the
  Cambridge Philosophical Society}}},\ Vol.~\bibinfo {volume} {34}\ (\bibinfo
  {organization} {Cambridge Univ Press},\ \bibinfo {year} {1938})\ pp.\
  \bibinfo {pages} {100--108}\BibitemShut {NoStop}%
\bibitem [{\citenamefont {Sondheimer}(1952)}]{sondheimer1952mean}%
  \BibitemOpen
  \bibfield  {author} {\bibinfo {author} {\bibfnamefont {E.~H.}\ \bibnamefont
  {Sondheimer}},\ }\bibfield  {title} {\enquote {\bibinfo {title} {The mean
  free path of electrons in metals},}\ }\href@noop {} {\bibfield  {journal}
  {\bibinfo  {journal} {Advances in physics}\ }\textbf {\bibinfo {volume}
  {1}},\ \bibinfo {pages} {1--42} (\bibinfo {year} {1952})}\BibitemShut
  {NoStop}%
\bibitem [{\citenamefont {Mayadas}\ and\ \citenamefont
  {Shatzkes}(1970)}]{mayadas1970electrical}%
  \BibitemOpen
  \bibfield  {author} {\bibinfo {author} {\bibfnamefont {A.}~\bibnamefont
  {Mayadas}}\ and\ \bibinfo {author} {\bibfnamefont {M.}~\bibnamefont
  {Shatzkes}},\ }\bibfield  {title} {\enquote {\bibinfo {title}
  {Electrical-resistivity model for polycrystalline films: the case of
  arbitrary reflection at external surfaces},}\ }\href@noop {} {\bibfield
  {journal} {\bibinfo  {journal} {Physical Review B}\ }\textbf {\bibinfo
  {volume} {1}},\ \bibinfo {pages} {1382} (\bibinfo {year} {1970})}\BibitemShut
  {NoStop}%
\bibitem [{\citenamefont {Graham}\ \emph {et~al.}(2010)\citenamefont {Graham},
  \citenamefont {Alers}, \citenamefont {Mountsier}, \citenamefont {Shamma},
  \citenamefont {Dhuey}, \citenamefont {Cabrini}, \citenamefont {Geiss},
  \citenamefont {Read},\ and\ \citenamefont {Peddeti}}]{graham2010resistivity}%
  \BibitemOpen
  \bibfield  {author} {\bibinfo {author} {\bibfnamefont {R.}~\bibnamefont
  {Graham}}, \bibinfo {author} {\bibfnamefont {G.}~\bibnamefont {Alers}},
  \bibinfo {author} {\bibfnamefont {T.}~\bibnamefont {Mountsier}}, \bibinfo
  {author} {\bibfnamefont {N.}~\bibnamefont {Shamma}}, \bibinfo {author}
  {\bibfnamefont {S.}~\bibnamefont {Dhuey}}, \bibinfo {author} {\bibfnamefont
  {S.}~\bibnamefont {Cabrini}}, \bibinfo {author} {\bibfnamefont
  {R.}~\bibnamefont {Geiss}}, \bibinfo {author} {\bibfnamefont
  {D.}~\bibnamefont {Read}}, \ and\ \bibinfo {author} {\bibfnamefont
  {S.}~\bibnamefont {Peddeti}},\ }\bibfield  {title} {\enquote {\bibinfo
  {title} {Resistivity dominated by surface scattering in sub-50 nm cu
  wires},}\ }\href@noop {} {\bibfield  {journal} {\bibinfo  {journal} {Applied
  Physics Letters}\ }\textbf {\bibinfo {volume} {96}},\ \bibinfo {pages}
  {042116} (\bibinfo {year} {2010})}\BibitemShut {NoStop}%
\bibitem [{\citenamefont {Steinh{\"o}gl}\ \emph {et~al.}(2005)\citenamefont
  {Steinh{\"o}gl}, \citenamefont {Schindler}, \citenamefont {Steinlesberger},
  \citenamefont {Traving},\ and\ \citenamefont
  {Engelhardt}}]{steinhogl2005comprehensive}%
  \BibitemOpen
  \bibfield  {author} {\bibinfo {author} {\bibfnamefont {W.}~\bibnamefont
  {Steinh{\"o}gl}}, \bibinfo {author} {\bibfnamefont {G.}~\bibnamefont
  {Schindler}}, \bibinfo {author} {\bibfnamefont {G.}~\bibnamefont
  {Steinlesberger}}, \bibinfo {author} {\bibfnamefont {M.}~\bibnamefont
  {Traving}}, \ and\ \bibinfo {author} {\bibfnamefont {M.}~\bibnamefont
  {Engelhardt}},\ }\bibfield  {title} {\enquote {\bibinfo {title}
  {Comprehensive study of the resistivity of copper wires with lateral
  dimensions of 100 nm and smaller},}\ }\href@noop {} {\bibfield  {journal}
  {\bibinfo  {journal} {Journal of Applied Physics}\ }\textbf {\bibinfo
  {volume} {97}},\ \bibinfo {pages} {023706} (\bibinfo {year}
  {2005})}\BibitemShut {NoStop}%
\bibitem [{\citenamefont {Feldman}\ \emph {et~al.}(2010)\citenamefont
  {Feldman}, \citenamefont {Park}, \citenamefont {Haverty}, \citenamefont
  {Shankar},\ and\ \citenamefont {Dunham}}]{feldman2010simulation}%
  \BibitemOpen
  \bibfield  {author} {\bibinfo {author} {\bibfnamefont {B.}~\bibnamefont
  {Feldman}}, \bibinfo {author} {\bibfnamefont {S.}~\bibnamefont {Park}},
  \bibinfo {author} {\bibfnamefont {M.}~\bibnamefont {Haverty}}, \bibinfo
  {author} {\bibfnamefont {S.}~\bibnamefont {Shankar}}, \ and\ \bibinfo
  {author} {\bibfnamefont {S.~T.}\ \bibnamefont {Dunham}},\ }\bibfield  {title}
  {\enquote {\bibinfo {title} {Simulation of grain boundary effects on
  electronic transport in metals, and detailed causes of scattering},}\
  }\href@noop {} {\bibfield  {journal} {\bibinfo  {journal} {physica status
  solidi (b)}\ }\textbf {\bibinfo {volume} {247}},\ \bibinfo {pages}
  {1791--1796} (\bibinfo {year} {2010})}\BibitemShut {NoStop}%
\bibitem [{\citenamefont {Ke}\ \emph {et~al.}(2009)\citenamefont {Ke},
  \citenamefont {Zahid}, \citenamefont {Timoshevskii}, \citenamefont {Xia},
  \citenamefont {Gall},\ and\ \citenamefont {Guo}}]{ke2009resistivity}%
  \BibitemOpen
  \bibfield  {author} {\bibinfo {author} {\bibfnamefont {Y.}~\bibnamefont
  {Ke}}, \bibinfo {author} {\bibfnamefont {F.}~\bibnamefont {Zahid}}, \bibinfo
  {author} {\bibfnamefont {V.}~\bibnamefont {Timoshevskii}}, \bibinfo {author}
  {\bibfnamefont {K.}~\bibnamefont {Xia}}, \bibinfo {author} {\bibfnamefont
  {D.}~\bibnamefont {Gall}}, \ and\ \bibinfo {author} {\bibfnamefont
  {H.}~\bibnamefont {Guo}},\ }\bibfield  {title} {\enquote {\bibinfo {title}
  {Resistivity of thin cu films with surface roughness},}\ }\href@noop {}
  {\bibfield  {journal} {\bibinfo  {journal} {Physical Review B}\ }\textbf
  {\bibinfo {volume} {79}},\ \bibinfo {pages} {155406} (\bibinfo {year}
  {2009})}\BibitemShut {NoStop}%
\bibitem [{\citenamefont {Timoshevskii}\ \emph {et~al.}(2008)\citenamefont
  {Timoshevskii}, \citenamefont {Ke}, \citenamefont {Guo},\ and\ \citenamefont
  {Gall}}]{timoshevskii2008influence}%
  \BibitemOpen
  \bibfield  {author} {\bibinfo {author} {\bibfnamefont {V.}~\bibnamefont
  {Timoshevskii}}, \bibinfo {author} {\bibfnamefont {Y.}~\bibnamefont {Ke}},
  \bibinfo {author} {\bibfnamefont {H.}~\bibnamefont {Guo}}, \ and\ \bibinfo
  {author} {\bibfnamefont {D.}~\bibnamefont {Gall}},\ }\bibfield  {title}
  {\enquote {\bibinfo {title} {The influence of surface roughness on electrical
  conductance of thin cu films: an ab initio study},}\ }\href@noop {}
  {\bibfield  {journal} {\bibinfo  {journal} {Journal of Applied Physics}\
  }\textbf {\bibinfo {volume} {103}},\ \bibinfo {pages} {113705} (\bibinfo
  {year} {2008})}\BibitemShut {NoStop}%
\bibitem [{\citenamefont {Zhou}\ \emph {et~al.}(2010)\citenamefont {Zhou},
  \citenamefont {Xu}, \citenamefont {Wang}, \citenamefont {Zhou},\ and\
  \citenamefont {Xia}}]{zhou2010ab}%
  \BibitemOpen
  \bibfield  {author} {\bibinfo {author} {\bibfnamefont {B.-h.}\ \bibnamefont
  {Zhou}}, \bibinfo {author} {\bibfnamefont {Y.}~\bibnamefont {Xu}}, \bibinfo
  {author} {\bibfnamefont {S.}~\bibnamefont {Wang}}, \bibinfo {author}
  {\bibfnamefont {G.}~\bibnamefont {Zhou}}, \ and\ \bibinfo {author}
  {\bibfnamefont {K.}~\bibnamefont {Xia}},\ }\bibfield  {title} {\enquote
  {\bibinfo {title} {An ab initio investigation on boundary resistance for
  metallic grains},}\ }\href@noop {} {\bibfield  {journal} {\bibinfo  {journal}
  {Solid State Communications}\ }\textbf {\bibinfo {volume} {150}},\ \bibinfo
  {pages} {1422--1424} (\bibinfo {year} {2010})}\BibitemShut {NoStop}%
\bibitem [{\citenamefont {C{\'e}sar}\ \emph {et~al.}(2014)\citenamefont
  {C{\'e}sar}, \citenamefont {Liu}, \citenamefont {Gall},\ and\ \citenamefont
  {Guo}}]{cesar2014calculated}%
  \BibitemOpen
  \bibfield  {author} {\bibinfo {author} {\bibfnamefont {M.}~\bibnamefont
  {C{\'e}sar}}, \bibinfo {author} {\bibfnamefont {D.}~\bibnamefont {Liu}},
  \bibinfo {author} {\bibfnamefont {D.}~\bibnamefont {Gall}}, \ and\ \bibinfo
  {author} {\bibfnamefont {H.}~\bibnamefont {Guo}},\ }\bibfield  {title}
  {\enquote {\bibinfo {title} {Calculated resistances of single grain
  boundaries in copper},}\ }\href@noop {} {\bibfield  {journal} {\bibinfo
  {journal} {Physical Review Applied}\ }\textbf {\bibinfo {volume} {2}},\
  \bibinfo {pages} {044007} (\bibinfo {year} {2014})}\BibitemShut {NoStop}%
\bibitem [{wil(2013)}]{wilson2013international}%
  \BibitemOpen
  \bibfield  {title} {\enquote {\bibinfo {title} {International technology
  roadmap for semiconductors (itrs), interconnects section},}\ }\href
  {http://www.itrs.net} {\bibfield  {journal} {\bibinfo  {journal}
  {Semiconductor Industry Association}\ } (\bibinfo {year} {2013})}\BibitemShut
  {NoStop}%
\bibitem [{\citenamefont {Soler}\ \emph {et~al.}(2002)\citenamefont {Soler},
  \citenamefont {Artacho}, \citenamefont {Gale}, \citenamefont {Garc{\'\i}a},
  \citenamefont {Junquera}, \citenamefont {Ordej{\'o}n},\ and\ \citenamefont
  {S{\'a}nchez-Portal}}]{soler2002siesta}%
  \BibitemOpen
  \bibfield  {author} {\bibinfo {author} {\bibfnamefont {J.~M.}\ \bibnamefont
  {Soler}}, \bibinfo {author} {\bibfnamefont {E.}~\bibnamefont {Artacho}},
  \bibinfo {author} {\bibfnamefont {J.~D.}\ \bibnamefont {Gale}}, \bibinfo
  {author} {\bibfnamefont {A.}~\bibnamefont {Garc{\'\i}a}}, \bibinfo {author}
  {\bibfnamefont {J.}~\bibnamefont {Junquera}}, \bibinfo {author}
  {\bibfnamefont {P.}~\bibnamefont {Ordej{\'o}n}}, \ and\ \bibinfo {author}
  {\bibfnamefont {D.}~\bibnamefont {S{\'a}nchez-Portal}},\ }\bibfield  {title}
  {\enquote {\bibinfo {title} {The siesta method for ab initio order-n
  materials simulation},}\ }\href@noop {} {\bibfield  {journal} {\bibinfo
  {journal} {Journal of Physics: Condensed Matter}\ }\textbf {\bibinfo {volume}
  {14}},\ \bibinfo {pages} {2745} (\bibinfo {year} {2002})}\BibitemShut
  {NoStop}%
\bibitem [{\citenamefont {Hegde}\ \emph
  {et~al.}(2014{\natexlab{a}})\citenamefont {Hegde}, \citenamefont
  {Povolotskyi}, \citenamefont {Kubis}, \citenamefont {Boykin},\ and\
  \citenamefont {Klimeck}}]{CuModel}%
  \BibitemOpen
  \bibfield  {author} {\bibinfo {author} {\bibfnamefont {G.}~\bibnamefont
  {Hegde}}, \bibinfo {author} {\bibfnamefont {M.}~\bibnamefont {Povolotskyi}},
  \bibinfo {author} {\bibfnamefont {T.}~\bibnamefont {Kubis}}, \bibinfo
  {author} {\bibfnamefont {T.}~\bibnamefont {Boykin}}, \ and\ \bibinfo {author}
  {\bibfnamefont {G.}~\bibnamefont {Klimeck}},\ }\bibfield  {title} {\enquote
  {\bibinfo {title} {An environment-dependent semi-empirical tight binding
  model suitable for electron transport in bulk metals, metal alloys, metallic
  interfaces, and metallic nanostructures. i. model and validation},}\ }\href
  {\doibase http://dx.doi.org/10.1063/1.4868977} {\bibfield  {journal}
  {\bibinfo  {journal} {Journal of Applied Physics}\ }\textbf {\bibinfo
  {volume} {115}},\ \bibinfo {eid} {123703} (\bibinfo {year}
  {2014}{\natexlab{a}})}\BibitemShut {NoStop}%
\bibitem [{\citenamefont {Hegde}\ \emph
  {et~al.}(2014{\natexlab{b}})\citenamefont {Hegde}, \citenamefont
  {Povolotskyi}, \citenamefont {Kubis}, \citenamefont {Charles},\ and\
  \citenamefont {Klimeck}}]{CuConfinement}%
  \BibitemOpen
  \bibfield  {author} {\bibinfo {author} {\bibfnamefont {G.}~\bibnamefont
  {Hegde}}, \bibinfo {author} {\bibfnamefont {M.}~\bibnamefont {Povolotskyi}},
  \bibinfo {author} {\bibfnamefont {T.}~\bibnamefont {Kubis}}, \bibinfo
  {author} {\bibfnamefont {J.}~\bibnamefont {Charles}}, \ and\ \bibinfo
  {author} {\bibfnamefont {G.}~\bibnamefont {Klimeck}},\ }\bibfield  {title}
  {\enquote {\bibinfo {title} {An environment-dependent semi-empirical tight
  binding model suitable for electron transport in bulk metals, metal alloys,
  metallic interfaces, and metallic nanostructures. ii. application—effect of
  quantum confinement and homogeneous strain on cu conductance},}\ }\href
  {\doibase http://dx.doi.org/10.1063/1.4868979} {\bibfield  {journal}
  {\bibinfo  {journal} {Journal of Applied Physics}\ }\textbf {\bibinfo
  {volume} {115}},\ \bibinfo {eid} {123704} (\bibinfo {year}
  {2014}{\natexlab{b}})}\BibitemShut {NoStop}%
\bibitem [{\citenamefont {QuantumWise}(2014)}]{atk}%
  \BibitemOpen
  \bibfield  {author} {\bibinfo {author} {\bibfnamefont {Q.~A.}\ \bibnamefont
  {QuantumWise}},\ }\href {http://www.quantumwise.com} {\enquote {\bibinfo
  {title} {Atomistix toolkit version 14.rc1},}\ } (\bibinfo {year}
  {2014})\BibitemShut {NoStop}%
\bibitem [{\citenamefont {Perdew}\ and\ \citenamefont
  {Zunger}(1981)}]{perdew1981self}%
  \BibitemOpen
  \bibfield  {author} {\bibinfo {author} {\bibfnamefont {J.~P.}\ \bibnamefont
  {Perdew}}\ and\ \bibinfo {author} {\bibfnamefont {A.}~\bibnamefont
  {Zunger}},\ }\bibfield  {title} {\enquote {\bibinfo {title} {Self-interaction
  correction to density-functional approximations for many-electron systems},}\
  }\href@noop {} {\bibfield  {journal} {\bibinfo  {journal} {Physical Review
  B}\ }\textbf {\bibinfo {volume} {23}},\ \bibinfo {pages} {5048} (\bibinfo
  {year} {1981})}\BibitemShut {NoStop}%
\bibitem [{\citenamefont {Datta}(2005)}]{datta2005quantum}%
  \BibitemOpen
  \bibfield  {author} {\bibinfo {author} {\bibfnamefont {S.}~\bibnamefont
  {Datta}},\ }\href@noop {} {\emph {\bibinfo {title} {Quantum transport: atom
  to transistor}}}\ (\bibinfo  {publisher} {Cambridge University Press},\
  \bibinfo {year} {2005})\BibitemShut {NoStop}%
\bibitem [{\citenamefont {Brandbyge}\ \emph {et~al.}(2002)\citenamefont
  {Brandbyge}, \citenamefont {Mozos}, \citenamefont {Ordej{\'o}n},
  \citenamefont {Taylor},\ and\ \citenamefont
  {Stokbro}}]{brandbyge2002density}%
  \BibitemOpen
  \bibfield  {author} {\bibinfo {author} {\bibfnamefont {M.}~\bibnamefont
  {Brandbyge}}, \bibinfo {author} {\bibfnamefont {J.-L.}\ \bibnamefont
  {Mozos}}, \bibinfo {author} {\bibfnamefont {P.}~\bibnamefont {Ordej{\'o}n}},
  \bibinfo {author} {\bibfnamefont {J.}~\bibnamefont {Taylor}}, \ and\ \bibinfo
  {author} {\bibfnamefont {K.}~\bibnamefont {Stokbro}},\ }\bibfield  {title}
  {\enquote {\bibinfo {title} {Density-functional method for nonequilibrium
  electron transport},}\ }\href@noop {} {\bibfield  {journal} {\bibinfo
  {journal} {Physical Review B}\ }\textbf {\bibinfo {volume} {65}},\ \bibinfo
  {pages} {165401} (\bibinfo {year} {2002})}\BibitemShut {NoStop}%
\bibitem [{\citenamefont {Kim}\ \emph {et~al.}(2010)\citenamefont {Kim},
  \citenamefont {Zhang}, \citenamefont {Nicholson}, \citenamefont {Evans},
  \citenamefont {Kulkarni}, \citenamefont {Radhakrishnan}, \citenamefont
  {Kenik},\ and\ \citenamefont {Li}}]{kim2010large}%
  \BibitemOpen
  \bibfield  {author} {\bibinfo {author} {\bibfnamefont {T.-H.}\ \bibnamefont
  {Kim}}, \bibinfo {author} {\bibfnamefont {X.-G.}\ \bibnamefont {Zhang}},
  \bibinfo {author} {\bibfnamefont {D.~M.}\ \bibnamefont {Nicholson}}, \bibinfo
  {author} {\bibfnamefont {B.~M.}\ \bibnamefont {Evans}}, \bibinfo {author}
  {\bibfnamefont {N.~S.}\ \bibnamefont {Kulkarni}}, \bibinfo {author}
  {\bibfnamefont {B.}~\bibnamefont {Radhakrishnan}}, \bibinfo {author}
  {\bibfnamefont {E.~A.}\ \bibnamefont {Kenik}}, \ and\ \bibinfo {author}
  {\bibfnamefont {A.-P.}\ \bibnamefont {Li}},\ }\bibfield  {title} {\enquote
  {\bibinfo {title} {Large discrete resistance jump at grain boundary in copper
  nanowire},}\ }\href@noop {} {\bibfield  {journal} {\bibinfo  {journal} {Nano
  letters}\ }\textbf {\bibinfo {volume} {10}},\ \bibinfo {pages} {3096--3100}
  (\bibinfo {year} {2010})}\BibitemShut {NoStop}%
\end{thebibliography}%

\end{document}